\documentclass[a4paper]{article}

\usepackage{INTERSPEECH2021}

\usepackage{multirow}

\usepackage[noadjust]{cite} 

\usepackage{relsize}

\usepackage[a-1b]{pdfx}

\title{Hierarchical Context-Aware Transformers\\for Non-Autoregressive Text to Speech}
\name{Jae-Sung Bae, Tae-Jun Bak, Young-Sun Joo, Hoon-Young Cho}
\address{Speech AI Lab, NCSOFT, Seongnam, Republic of Korea}
\email{\{jaesungbae, happyjun, ysjoo555, hycho\}@ncsoft.com}

\begin{document}

\maketitle
\begin{abstract}
In this paper, we propose methods for improving the modeling performance of a Transformer-based non-autoregressive text-to-speech (TNA-TTS) model. Although the text encoder and audio decoder handle different types and lengths of data (i.e., text and audio), the TNA-TTS models are not designed considering these variations. Therefore, to improve the modeling performance of the TNA-TTS model we propose a hierarchical Transformer structure-based text encoder and audio decoder that are designed to accommodate the characteristics of each module. For the text encoder, we constrain each self-attention layer so the encoder focuses on a text sequence from the local to the global scope. Conversely, the audio decoder constrains its self-attention layers to focus in the reverse direction, i.e., from global to local scope. Additionally, we further improve the pitch modeling accuracy of the audio decoder by providing sentence and word-level pitch as conditions. Various objective and subjective evaluations verified that the proposed method outperformed the baseline TNA-TTS.
\end{abstract}

\section{Introduction}
\label{sec:intro}
Recently, the naturalness of synthesized speech and the controllability of text-to-speech (TTS) models have been significantly improved due to the advancements in neural TTS models \cite{tacotron2, deepvoice3, gst, jsbae}. Transformer-based non-autoregressive TTS (TNA-TTS) models consisting of two feed-forward Transformer (FFT) \cite{transformer} stacks, i.e., the encoder and decoder, have been proposed \cite{fastspeech, fastpitch, fastspeech2, aligntts}. Compared to sequence-to-sequence autoregressive TTS \cite{tacotron2, deepvoice3}, these models not only speed up the inference speed but also enhance the robustness of synthesized speech with significantly less skipping, repetition, and deletion.

FastPitch \cite{fastpitch} and FastSpeech2 \cite{fastspeech2} ease the one-to-many problem of TTS by providing additional acoustic features, such as pitch or energy sequences, to the decoder. By modifying the acoustic features, synthesized speech could be controlled. However, because speech components are correlated to each other, inaccurately predicted feature values affect the final synthesized speech quality. In other words, as the number of additional input features increases, the performance dependency of the TTS model on these features also increases. Therefore, there is still a performance gap that must be improved.

The self-attention mechanism is a core component of Transformers that helps capture the contextual information from the entire sequence and enables Transformers to demonstrate remarkable performance in various domains \cite{transformerTTS, bert, music_transformer1}. However, this mechanism might prevent networks from focusing on the local information that is also important in many cases. For this reason, various studies that constrain the attention range of Transformers have been investigated. In the natural language processing (NLP) field, constraining the attention range to focus on local context has improved the modeling performance in long sequence tasks \cite{sparse, adaptive_attn_span, longformer, bigbird}. Based on these studies, we expect such approaches would also be effective in the text encoder of TTS models. In \cite{gmvae, icassp2020_fullyhie, hierarchical_multi_grained, chive}, TTS models with a hierarchical architecture have been studied, and it was reported that focusing from the overall part of the input sequence to the detailed part was effective for the audio decoder. 

In our preliminary experiments, we analyzed the average attention weights of the text encoder and audio decoder in a TNA-TTS model; FastPitch \cite{fastpitch} was used as the TNA-TTS model. We did not observe any specific tendencies for all layers in the attention weights of the encoder. The decoder focused on the local context at the layers close to the input, whereas it focused on the global context at the layers close to the output. This is a contrary tendency to the prior studies on the hierarchical TTS architecture. Further details about the experiments are discussed in Section \ref{sec:res_analysis_attn_weight}.

To improve the modeling performance of TNA-TTS, we propose hierarchical Transformer structures for the text encoder and the audio decoder considering their characteristics. For the hierarchical Transformer structure, we constrain the attention scope of each self-attention layer of the encoder and decoder by applying a \emph{windowed attention} pattern with a fixed-size window. The windowed attention pattern makes the self-attention layer focus on data within the scope. By modifying the window size, we control the focusing scope. For the text encoder, we set the window size progressively narrower as the layer gets closer to the input text embedding. In other words, the text encoder of the proposed structure focuses on the local context of the text embedding at first, and as the network deepens it focuses on the overall context. Additionally, we apply a \emph{global attention} pattern \cite{longformer, bigbird} for the encoder, so that the text symbols in the sentences, such as exclamation points and question marks, can be emphasized. The global attention pattern is exceptionally accessible to anywhere in the input sequence regardless of the attention constraints. For the audio decoder, in the opposite direction, we constrain the attention to have a wider scope closer to the decoder input layer, which progressively narrows as the layers approach the decoder output. That is, the audio decoder initially generates a rough mel-spectrogram, then refines the details of the mel-spectrogram by focusing on its local part as the layer goes up.

Because the pitch continuously changes over time, the audio decoder of the FastPitch implicitly predicts a detailed pitch contour from the character-level pitch embedding inputs and generates a mel-spectrogram containing the pitch contour. In \cite{f0-modeling}, long-term fundamental frequency ($F_0$) features improved the accuracy of $F_0$ modeling. Therefore, to further improve the modeling performance of the audio decoder, we additionally provide the hierarchical pitch embeddings, which are sentence and word-level pitch embeddings, to the audio decoder.

\section{Background}

\subsection{Self-Attention}

The self-attention layer is a core component of the Transformer \cite{transformer} that allows the Transformer to explicitly calculate the relationship between input sequences and capture the long-term dependencies. In the self-attention layer, the same input vector is used as query, key, and value of the attention mechanism. The Transformer adopts a scaled-dot product attention mechanism for the self-attention layer. It is defined as follows: 

\begin{equation}
\mathrm{Score}(Q,K)=\frac{(QW^Q)(KW^K)^T}{\sqrt{d}},
\label{transformer1}
\end{equation}
\begin{equation}
\text{Attention}(Q,K,V)=\sigma \left( \mathrm{Score}(Q,K) \right)(VW^V),
\label{transformer2}
\end{equation}
where $\sigma(\cdot)$ denotes the softmax operation.
$Q,K,V\in\mathbb{R}^{n\times d}$ are matrices of $n$ query, key, and value vectors, and $W^Q,W^K,W^V\in\mathbb{R}^{d\times d}$ are the corresponding weights of the linear projection layer. $d$ denotes a dimension of the vector. First, an attention score between $Q$ and $K$ is calculated, then it is passed through the softmax function. The output of the softmax function is referred to as an \textit{attention weight}. To compute the final output of the self-attention layer, the attention weight is multiplied by the linear projected input $V$.

\subsection{Attention Patterns}
\label{sec:attention_pattern}
\begin{figure}[t]
{\small
    \begin{minipage}[b]{.32\linewidth}
      \centering
      \centerline{\includegraphics[width=2.4cm]{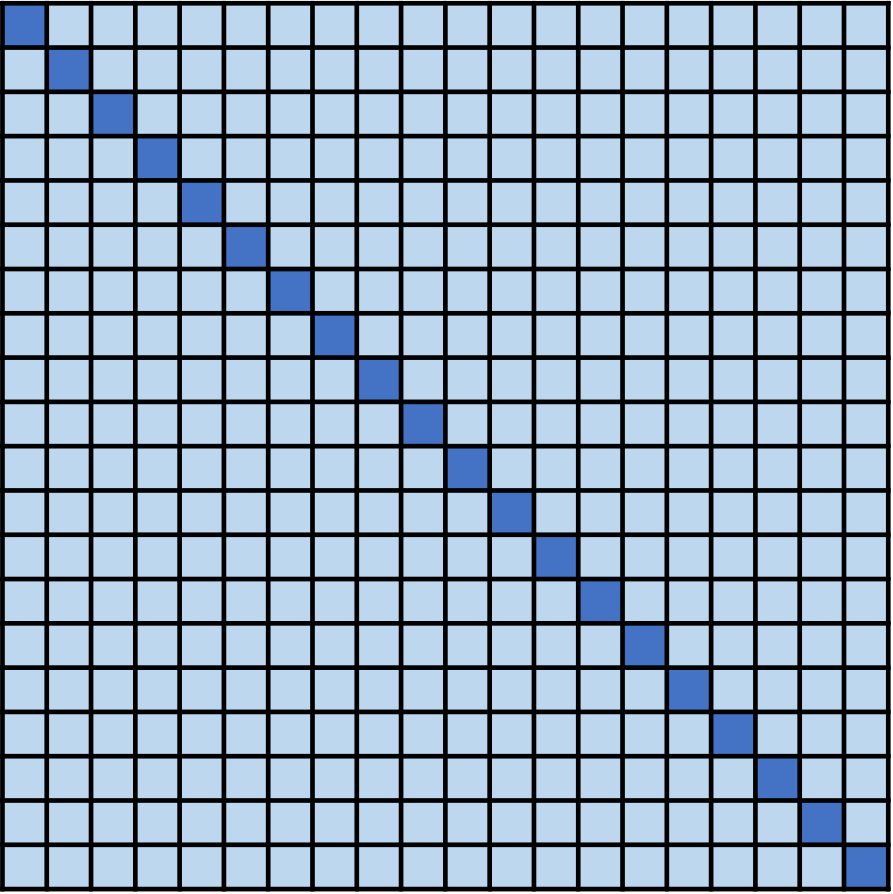}}
      \centerline{(a)}
    \end{minipage}
    \begin{minipage}[b]{.32\linewidth}
      \centering
      \centerline{\includegraphics[width=2.4cm]{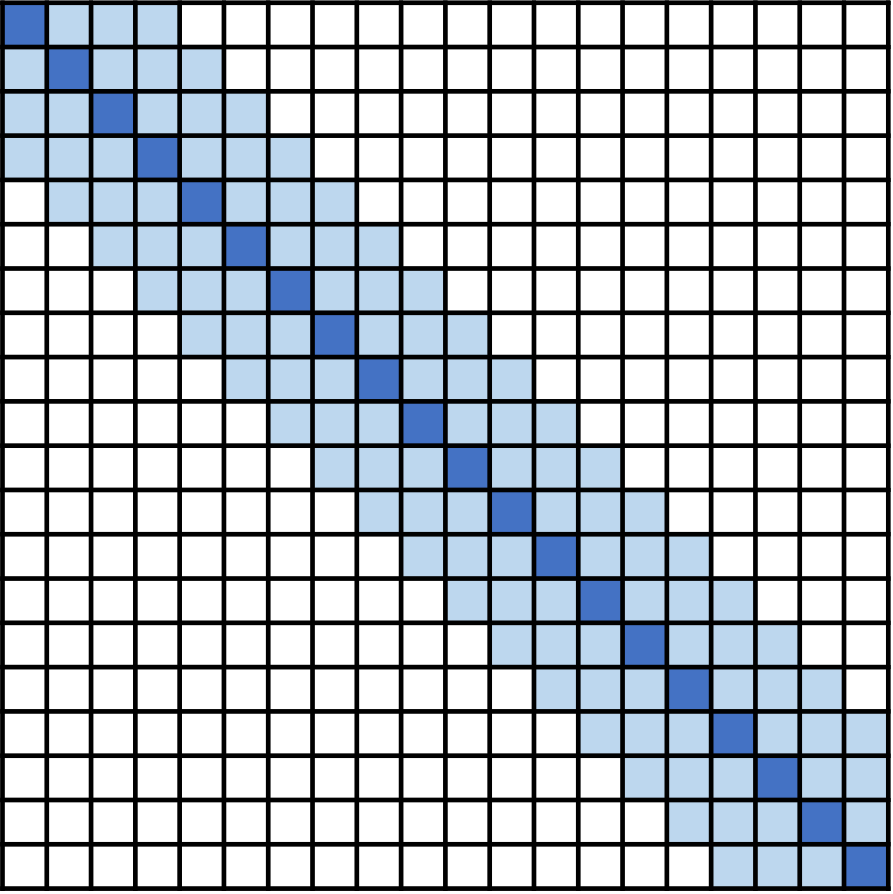}}
      \centerline{(b)}
    \end{minipage}
    \begin{minipage}[b]{.32\linewidth}
      \centering
      \centerline{\includegraphics[width=2.4cm]{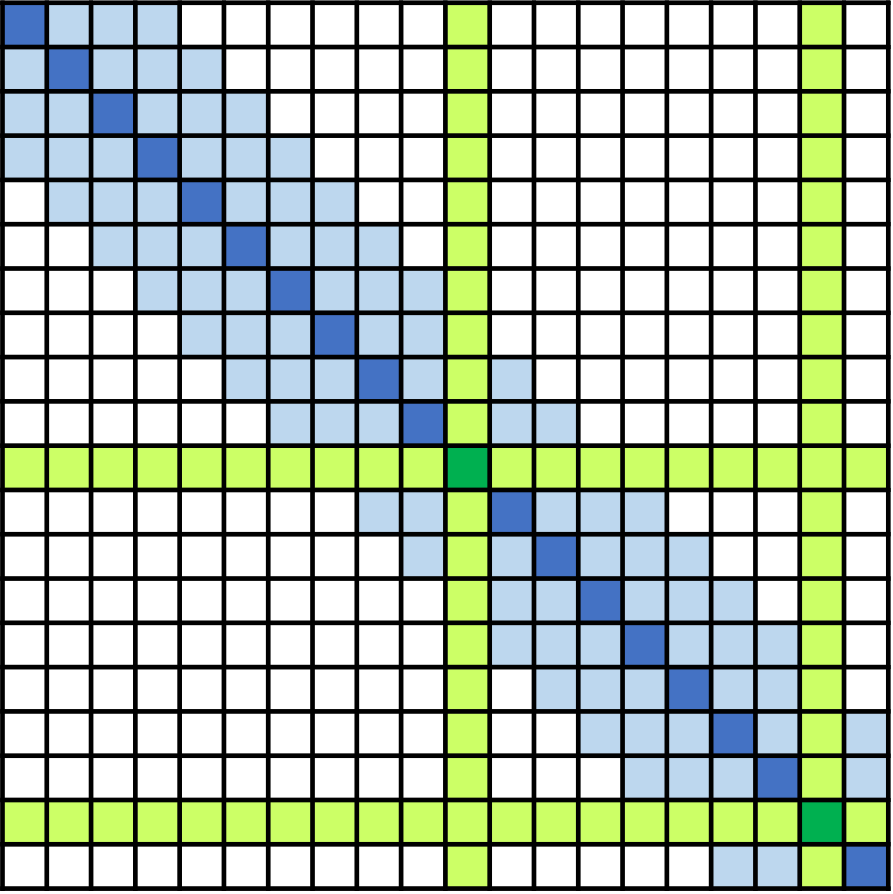}}
      \centerline{(c)}
    \end{minipage}
    \caption{Example of attention patterns. (a) full attention, (b) windowed attention with $w=6$, and (c) combination of windowed attention (blue) and global attention on a few pre-selected locations (green). The column and row of matrices are the position of query and key vectors, respectively.}
    \label{fig:attn}
}
\end{figure}
In \cite{sparse, adaptive_attn_span, longformer, bigbird}, the attention range was constrained through the \textit{attention patterns} to make the network focus on proper information. The attention pattern is element-wise multiplied to the attention score. In the following, we explain two types of patterns, \textit{windowed attention} and \textit{global attention}, used in our proposed hierarchical Transformer structure. 
\\

\noindent
\textbf{Windowed Attention: } This is an attention pattern with a fixed window size of $w$, as depicted in Figure \ref{fig:attn}b. It calculates attention only within a window and does not calculate attention for tokens outside the window range. It helps to focus on local information rather than the whole information. By adjusting the window size, the focusing scope can be controlled.
\\

\noindent
\textbf{Global Attention: } 
This is introduced to learn task-specific representations by adding it to pre-selected input locations, e.g., exclamation and question marks in a text sequence. As shown in Figure \ref{fig:attn}c, combined with the windowed attention, it exceptionally enables every input sequence of the attention layer to attend to specific tokens regardless of its attention constraints and also enables specific tokens to attend to every input sequence. This compensates for the disadvantage of inflexible windowed attention.

\section{Proposed Method}

\begin{figure}[t]
{\small
      \centering
      \centerline{\includegraphics[width=\linewidth]{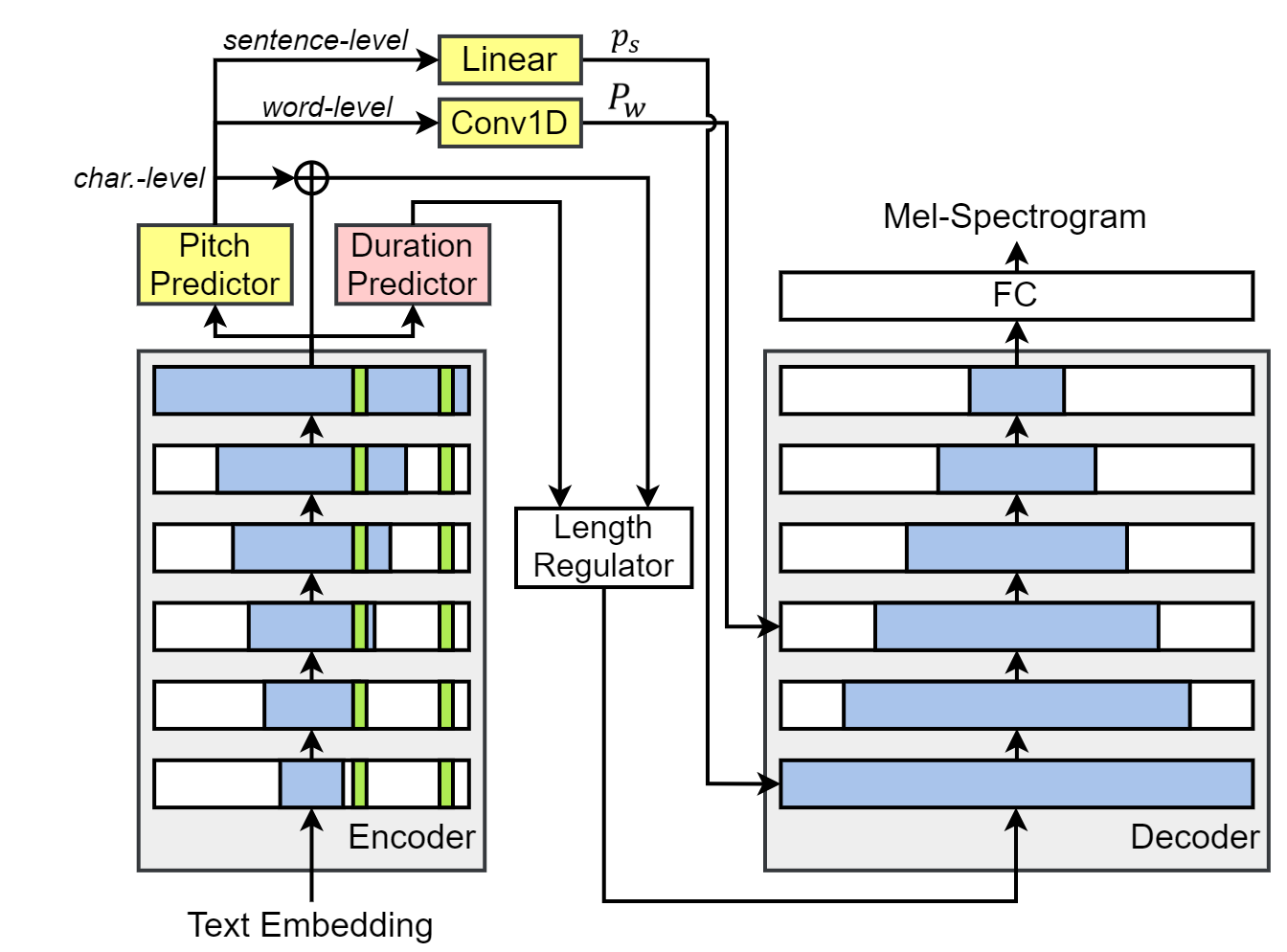}}
        \caption{Architecture of hierarchical context-aware Transformers applied to FastPitch \cite{fastpitch}. Each block in the encoder and decoder represents a single FFT layer. In each FFT layer, the blue box illustrates the windowed attention and its horizontal length means the window size. The green lines depict the global attentions on specific tokens. (Best viewed in color).}
    \label{fig:model}
}
\end{figure}

\label{sec:proposed}
Figure \ref{fig:model} illustrates the proposed architectures of the TNA-TTS model which applies the hierarchical Transformers structure. We apply the windowed attention to determine the scope of information to be focused on in each layer of the encoder and decoder. Further, we decide the window size considering the characteristics of the encoder and decoder. For further improving the modeling performance of the decoder, we provide sentence and word-level pitch embeddings to the decoder.

\subsection{Baseline TNA-TTS Model}
For the TNA-TTS model, we used FastPitch \cite{fastpitch} model. FastPitch consists of a text encoder, audio decoder, duration predictor, and pitch predictor. The encoder converts an input text embedding sequence into a hidden representation sequence. Based on the hidden representation sequence, character-level pitch and duration values are predicted. The pitch values are embedded into pitch embedding and the embedding is added to hidden representation. After the length-regulating process, it is finally decoded into a mel-spectrogram.

\subsection{Hierarchical Attention Pattern}
We designed the encoder and decoder architectures while considering the encoder and decoder characteristics. For the encoder, we set the window size of the lower layers, i.e. close to the input text sequence, to be small and increase the window size as the layer becomes deeper. In the final layer, full attention is applied. This enables important tokens to affect the overall sequence and improves the performance. In addition to the windowed attention, we apply the global attention to pre-selected symbols, e.g., exclamation or question marks, in text sequences. This is because such symbols in a text sequence are important factors that affect the prosody of the generated speech in TTS. 

For the decoder, the attention pattern of the first layer is full attention. Then, we set the window size of the lower layers to be large and decrease as the layer gets deeper. In the lower layer, it generates the overall structure of the mel-spectrogram by focusing on the global context of the input; then, in the upper layer, it refines the mel-spectrogram to have better intelligibility by focusing on the local context.

\subsection{Hierarchical Pitch Conditioning}
We provide sentence and word-level pitch embeddings as additional conditions for the lower attention layers to further improve the modeling performance of the decoder. Because the proposed decoder structure focuses on the global information rather than the local at the lower layers, the layer conditioned on the sentence-level pitch embedding is lower than that conditioned on the word-level pitch embedding. Providing additional pitch embeddings allows the decoder to observe the overall pitch contour as well as the details of the pitch contour.

The details of the procedures are as follows. For a sentence-level pitch embedding, first, all character-level pitch values of the input sentence are averaged as a single scalar value. Then, the value is embedded into a sentence-level pitch embedding vector $p_s\in\mathbb{R}^{d}$ with a single linear projection layer, where $d$ denotes the embedding dimension. For the word-level pitch embedding, first, the character-level pitch values are averaged for each word. Then, the word-level pitch values are embedded into $P_w\in\mathbb{R}^{n\times d}$ by using a single convolutional layer with a kernel size of 3 and a stride of 1; where $n$ denotes the number of words in the input sentence. The sentence and word-level pitch embeddings are replicated to match the decoder input sequence length, respectively, where the word-level duration calculated from the corresponding character-level duration is used. We provide these pitch embeddings to the corresponding decoder self-attention layers by conditioning them to the attention score as follows:
\begin{equation}
\mathrm{Score}(Q,K,P)=\frac{(QW^Q + P)(KW^K)^T}{\sqrt{d}},
\end{equation}
where $P\in\mathbb{R}^{t\times d}$ denotes a replicated pitch embedding and $t$ is the input sequence length of the first decoder layer.

\section{Experiments}
\label{sec:exp}
 
\subsection{Dataset} 
For the experiments, we used an internal Korean female speech dataset comprising 22 h of data with 11000 recorded sentences. 
The dataset comprised 22 h of data with 11000 recorded sentences. 
The speech sample length varied from 1.5–13 s. The dataset was recorded at a frequency sampling rate of 22050 Hz in a studio environment. Each one percent of the dataset was randomly selected and used as a test and validation set, respectively. The rest of the dataset was used for training. As the acoustic feature, an 80-bin mel-spectrogram with a hop size of 256 and window size of 1024 was computed. We extracted character-level duration values using a pre-trained Transformer TTS \cite{transformerTTS} with a single attention head; the same dataset was used for pre-training. The character-level pitch values were obtained by averaging the pitch values for each character. The PRAAT toolkit \cite{praat} was used to extract pitch values from recorded speech.

\subsection{Model Setup}
The architectures of the baseline TNA-TTS model, which is FastPitch, were the same as that described in the original paper \cite{fastpitch}. For windowed attention, the window size of the encoder from the first to the fifth layer was set to 10, 20, 40, 60, and 100; that of the decoder from the second to the last layer was set to 400, 200, 100, 60, and 40; these values were determined empirically. Full attention was used in the last encoder layer and first decoder layer. The sentence and word-level pitch embeddings were provided as conditions to the first and third layers of the decoder, respectively. The pitch embedding dimension, $d$, was set to 64, which is the same as that of self-attention.

FastPitch is generally trained under the warm-up learning rate schedule; however, we found that using a half-learning rate schedule for every 40000 iterations yielded better performance. The ratio of duration, pitch, and mel-spectrogram prediction loss was the same in the original paper; however, we obtained better performance when the ratio was 0.01:0.01:1. The Adam optimizer \cite{adam} with $\beta_1=0.5$, $\beta_2=0.9$, $\epsilon=10^{-6}$ was used to optimize the network with an initial learning rate of 0.002, and the batch size was 16. For a neural vocoder, VocGAN \cite{vocgan} was used. It was trained using a database containing approximately 40 h of speech recorded by six speakers.

\subsection{Analysis on Attention Weights in FastPitch}
\label{sec:res_analysis_attn_weight}

\setlength{\belowcaptionskip}{-10pt}
\begin{figure}[t]
{\small
      \centering
      \centerline{\includegraphics[width=0.9\linewidth]{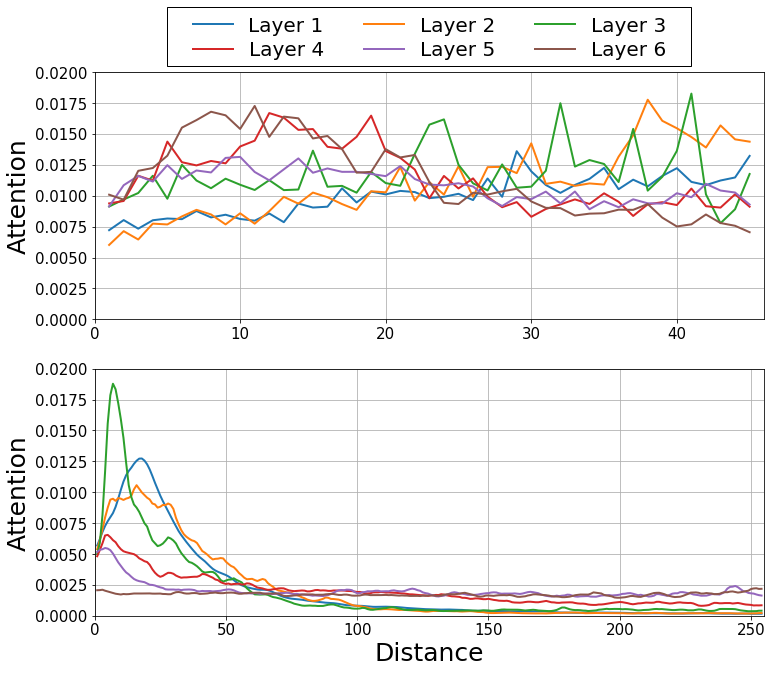}}
        \caption{Example of average attention weights of the baseline model encoder (top) and decoder (bottom) layers according to the location distance between the query and key vectors. Without any constraint, any tendency is not observed in the encoder. For the decoder, the lower layers focus on the local information while the upper layers focus on the global information. (Best viewed in color).}
    \label{fig:attn_result}
}
\end{figure}

In this subsection, we analyze the attention weights of the encoder and decoder in the baseline FastPitch. Figure \ref{fig:attn_result} depicts the average attention weights of each layer according to the location distance between the query and key vectors. Please note that the different distance ranges of the encoder and decoder are because of the different data types that they are handling, i.e., text and audio. A peak at the smaller distance implies the layer focuses on the local context. As illustrated in Figure \ref{fig:attn_result}, when there are no restrictions on the attention range, it is difficult to observe any tendency for the encoder. Whereas, the decoder focuses on the local context at the lower layers and the global context at the higher layers.

\subsection{Objective Evaluation}

\setlength{\tabcolsep}{4pt}
\begin{table}[t]
    {\small  
    \caption{Objective evaluation results with a confidence interval (CI) of 95\%. Lower is better for all metrics. GT denotes the ground-truth.}
    \label{res:total}
    \begin{center}      
    \begin{tabular}{lrrrr}
        \toprule
        \multicolumn{1}{l}{\textbf{Model}} &
        \multicolumn{1}{c}{\textbf{FFE (\%)}} & 
        \multicolumn{1}{c}{\textbf{MCD (dB)}} & 
        \multicolumn{1}{c}{\textbf{CER (\%)}} \\
        \midrule
        GT (recorded) & - & - & $4.96$ \\ 
        GTMel+Vocoder & $3.23\pm0.26$ & $2.26\pm0.02$ & $5.11$ \\
        \midrule
        FastPitch & $8.89\pm0.47$ & $6.59\pm0.18$ & $15.32$ \\
        \midrule
        EGW & $8.55\pm0.48$ & $6.54\pm0.17$ & $11.52$ \\
        DW & $8.21\pm0.45$ & $6.26\pm0.15$ & $10.58$ \\  
        EGW+DW & $8.13\pm0.43$ & $\mathbf{6.16\pm0.14}$ & $\mathbf{9.16}$\\
        EGW+DW+HPC & $\mathbf{8.00\pm0.45}$ & $\mathbf{6.16\pm0.13}$ & $9.88$
        \\
        \bottomrule
    \end{tabular}
    \end{center}
    }
\end{table}

For the experiments, we compared five models: FastPitch, FastPitch applying the global and windowed attention to the encoder (EGW), FastPitch applying the windowed attention to the decoder (DW), FastPitch with both EGW and DW (EGW+DW), and FastPitch applying EGW, DW, and the hierarchical pitch conditioning (EGW+DW+HPC). For objective evaluation, the f0 frame error (FFE) \cite{ffe}, mel-cepstral distance (MCD) \cite{mcd}, and character error rate (CER) were measured between the recorded speech and the generated speech samples.
In the case of FFE and MCD, we generated speech samples by using ground-truth values of the character-level pitch and duration to measure the encoder and decoder performance excluding the effect caused by inaccurate prediction of the pitch and duration predictors. For the CER, we evaluated the final synthesized speech which was generated with predicted pitch and duration values. The CER was measured by using a pre-trained Kaldi \cite{kaldi}-based automatic speech recognition model.

The results are summarized in Table \ref{res:total}. The proposed method significantly improved the overall performance of the FastPitch. The proposed hierarchical Transformer structure was effective not only when applied to both the encoder and decoder, but also when applied to the encoder and decoder, respectively. By additionally adopting HPC, it achieved the best performance for the FFE and MCD, even though slight degradation was observed on the CER.

\subsection{Ablation Study on Hierarchical Structure}

\setlength{\tabcolsep}{4pt}
\begin{table}[t]
{\small
    \caption{Ablation study results on the hierarchical structure with 95\% CIs.}
    \label{res:ab}
    \begin{center}      
    \begin{tabular}{lrrrr}
        \toprule
        \multicolumn{1}{l}{\textbf{Model}} &
        \multicolumn{1}{c}{\textbf{FFE (\%)}} & 
        \multicolumn{1}{c}{\textbf{MCD (dB)}} & 
        \multicolumn{1}{c}{\textbf{CER (\%)}} \\
        \midrule
        EGW (10 to full) & $\mathbf{8.55\pm0.48}$ & $\mathbf{6.54\pm0.17}$ & $\mathbf{11.52}$ \\
        EGW (full to 10) & $8.95\pm0.50$ & $\mathbf{6.54\pm0.17}$ & $14.66$ \\
        \midrule
        DW (40 to full) & $8.58\pm0.48$ & $6.42\pm0.18$ & $11.45$ \\
        DW (full to 40) & $\mathbf{8.21\pm0.45}$ & $\mathbf{6.26\pm0.15}$ & $\mathbf{10.58}$ \\  
        \midrule
        HPC & $8.96\pm0.52$ & $6.60\pm0.17$ & $13.49$\\
        \bottomrule
    \end{tabular}
    \end{center}
}
\end{table}

Here, we demonstrate the importance of considering the encoder and decoder characteristics in the proposed hierarchical Transformer structure via ablation study. Table \ref{res:ab} presents the ablation study results. For the encoder, the increasing window size yielded better performance than the decreasing window size, which is consistent with previous NLP task-related studies. 
Conversely, the increasing window size in the decoder performed worse than the decreasing window size. This implies that focusing on the context from the global to local scale, hierarchically, is important for predicting the mel-spectrogram. This is because the TTS decoder first generates the overall architecture of the mel-spectrogram in the lower layers, and then refines it. 

Finally, we evaluated the case wherein only the HPC was adopted. Without any constraint in the attention range, providing additional pitch information in the lower decoder layer did not produce any effects. This is because the baseline model decoder focuses on the local context in the lower layers; therefore, an appropriate hierarchical structure of the decoder is also important for the HPC to be effective.

\subsection{Subjective Evaluation}

\begin{table}[t]
{\small
    \caption{MOS test results for naturalness with 95\% CIs.}
    \label{res:mos}
    \begin{center}      
    \begin{tabular}{lrrrr}
        \toprule
        \multicolumn{1}{l}{\textbf{Model}} &
        \multicolumn{1}{c}{\textbf{MOS}} \\
        \midrule
        FastPitch (baseline) & $3.75 \pm 0.07$\\
        EGW+DW & $3.96 \pm 0.07$\\
        EGW+DW+HPC & $\mathbf{4.06 \pm 0.06}$ \\
        \bottomrule
    \end{tabular}
    \end{center}
}
\end{table}
To evaluate the naturalness of the generated speech, we conducted mean opinion score (MOS) tests. We compared the FastPitch, EGW+DW, and EGW+DW+HPC models. For each model, 20 samples were generated with texts included in the test set. A total of 31 native Koreans participated and scored the naturalness of the generated speech samples from 1 to 5.

The results\footnote{Audio samples can be found online: \\ https://nc-ai.github.io/speech/publications/hierarchical-transformers-tts} presented in Table \ref{res:mos} indicate that the EGW+\\DW+HPC model achieved the highest score, 4.06. Even though the EGW+DW received a slightly lower score of 3.96, it is still a significantly higher score compared to the baseline. FastPitch obtained a lower score because the synthesized speech contained ambiguous pronunciation. Please note that the difference in the dataset, language, and vocoder type used for the experiments might be the reasons for the lower MOS score of FastPitch than that reported in the original paper \cite{fastpitch}.

\section{Conclusion}
\label{sec:conclusion}
This paper proposed a hierarchical context-aware Transformers structure for the TNA-TTS model. Considering the different characteristics of the encoder and decoder of TNA-TTS, we constrained the attention range of the encoder and decoder differently. This enabled the encoder and decoder to focus on appropriate contextual information at each layer. To further improve the pitch modeling accuracy of the decoder, we provided hierarchical pitch conditioning, i.e., sentence and word-level pitch embedding, to the decoder. We verified the superior performance of the proposed methods through various quantitative and qualitative evaluations. In future work, instead of the fixed window size, a learnable window size for the windowed attention could be researched. This would help to find the optimal hierarchical Transformer structure for the encoder and decoder.

\pagebreak
\bibliographystyle{IEEEtran}
\bibliography{mybib}

\end{document}